\documentclass[aip,reprint]{revtex4-1}

\draft 

\usepackage{graphicx}
\usepackage{dcolumn}
\usepackage{bm}

\usepackage[utf8]{inputenc}
\usepackage[T1]{fontenc}
\usepackage{mathptmx}

\begin{document}


\title{Energy-Time Entanglement-based Dispersive Optics Quantum Key Distribution over Optical Fibers of 20 km} 

\author{Xu Liu}
 \affiliation{Beijing National Research Center for Information Science and Technology (BNRist), Beijing Innovation Center for Future Chips, Electronic Engineering Department, Tsinghua University, Beijing 100084, China.}
 \affiliation{Beijing Academy of Quantum Information Sciences, Beijing 100193, China.}
\author{Xin Yao}%
\affiliation{Beijing National Research Center for Information Science and Technology (BNRist), Beijing Innovation Center for Future Chips, Electronic Engineering Department, Tsinghua University, Beijing 100084, China.}
 \affiliation{Beijing Academy of Quantum Information Sciences, Beijing 100193, China.}

\author{Heqing Wang}%
\affiliation{State Key Laboratory of Functional Materials for Informatics, Shanghai Institute of Microsystem and Information Technology, Chinese Academy of Sciences, Shanghai 200050, China.}
\author{Hao Li}%
\affiliation{State Key Laboratory of Functional Materials for Informatics, Shanghai Institute of Microsystem and Information Technology, Chinese Academy of Sciences, Shanghai 200050, China.}

\author{Zhen Wang}%
\affiliation{State Key Laboratory of Functional Materials for Informatics, Shanghai Institute of Microsystem and Information Technology, Chinese Academy of Sciences, Shanghai 200050, China.}

\author{Lixing You}%
\affiliation{State Key Laboratory of Functional Materials for Informatics, Shanghai Institute of Microsystem and Information Technology, Chinese Academy of Sciences, Shanghai 200050, China.}

\author{Yidong Huang}%
 \affiliation{Beijing National Research Center for Information Science and Technology (BNRist), Beijing Innovation Center for Future Chips, Electronic Engineering Department, Tsinghua University, Beijing 100084, China.}
 \affiliation{Beijing Academy of Quantum Information Sciences, Beijing 100193, China.}

\author{Wei Zhang}%
\email{zwei@tsinghua.edu.cn}
 \affiliation{Beijing National Research Center for Information Science and Technology (BNRist), Beijing Innovation Center for Future Chips, Electronic Engineering Department, Tsinghua University, Beijing 100084, China.}
 \affiliation{Beijing Academy of Quantum Information Sciences, Beijing 100193, China.}

\date{\today}

\begin{abstract}
An energy-time entanglement-based dispersive optics quantum key distribution (DO-QKD) is demonstrated experimentally over optical fibers of 20 km. In the experiment, the telecom band energy-time entangled photon pairs are generated through spontaneous four wave mixing in a silicon waveguide. The arrival time of photons are registered for key generating and security test. High dimensional encoding in the arrival time of photons is used to increase the information per coincidence of photon pairs. The bin sifting process is optimized by a three level structure, which significantly reduces the raw quantum bit error rate (QBER) due to timing jitters of detectors and electronics. A raw key generation rate of 151kbps with QBER of 4.95\% is achieved, under a time-bin encoding format with 4 bits per coincidence. This experiment shows that entanglement-based DO-QKD can be implemented in an efficient and convenient way, which has great potential in quantum secure communication networks in the future. 
\end{abstract}

{03.67.Dd, 03.67.Hk}

\maketitle 
Quantum key distribution (QKD) permits remote parties to share secret keys. The secret keys can be further used for symmetric encryption, such as the one-time pad encryption scheme, which provides information-theoretic secure message transmission. Since the first protocol (BB84) \cite{bennett1984quantum} was proposed, QKD has been developed significantly \cite{PhysRevLett.68.557,PhysRevLett.94.230503,doi:10.1063/1.2432296,doi:10.1063/1.5027030}. From the point of QKD implementation, there are mainly two kinds of QKD schemes, which are the prepare-and-measurement (P\&M) schemes \cite{PhysRevLett.98.010505,Lucamarini:13} and the entanglement-based schemes \cite{PhysRevA.76.012307,doi:10.1063/1.2348775}. The entanglement-based QKD has inherent random local results and correlated outcomes. No extra random number generators are required in these schemes. On the other hand, if it could be swapped with high fidelity by quantum repeater nodes, photonic entanglement could be extended over long distances, which is promising to realize global scale quantum networks \cite{PhysRevLett.81.5932}. Hence, the entanglement-based QKD are important as a basic function of future quantum secure communication networks.

\begin{figure*}[htbp]
\setlength{\belowcaptionskip}{-0.2cm}  
\centering\includegraphics[scale=0.5]{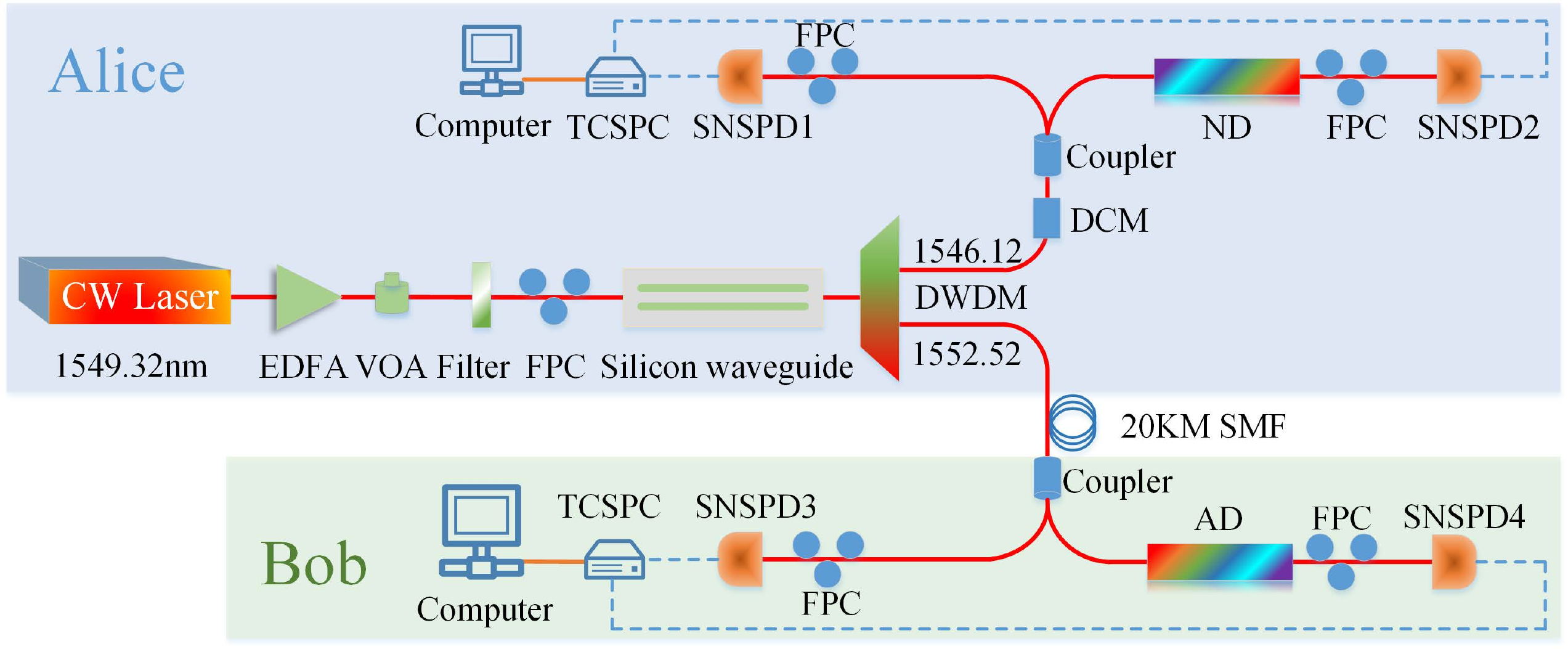}%
\caption{\label{fig1}The setup of entanglement-based DO-QKD. EDFA: Erbium doped fiber amplifier; VOA: variable optical attenuator; FPC: fiber polarization controller; DWDM: dense wavelength division multiplexing filter; DCM: dispersion compensation module; ND: normal dispersion; AD: anomalous dispersion; TCSPC: time-correlated single-photon counting module.}%
\end{figure*}

Dispersive optics QKD (DO-QKD) is a promising QKD protocol \cite{PhysRevA.87.062322} developed in recent years, which is robust in system stability and convenient in implementation. In DO-QKD, normal and anomalous dispersion components are introduced at Alice and Bob sides, constructing the time-frequency bases. The security test of DO-QKD can be realized by measurements of unbiased time-frequency bases, which has been proven to be secure against collective attacks \cite{PhysRevA.87.062322}. The P\&M protocols of DO-QKD has been demonstrated by field experiments with a high key rate of 1.2 Mbps over optical fibers of 43km \cite{lee2016high}. However, the entanglement-based DO-QKD was only demonstrated by short fiber links with relatively low key generation rates \cite{PhysRevA.90.062331}. In this work, we experimentally realized the entanglement-based DO-QKD over single mode optical fibers of 20 km. Energy-time entangled photon pairs were used in this experiment, which were generated by spontaneous four wave mixing (SFWM) in piece of silicon waveguide. In order to enhance the key generation rate, high dimensional encoding was used to increase the information per coincidence of photon pairs. On the other hand, the bin sifting process was optimized by a three level structure, which significantly reduced the quantum bit error rate (QBER) of raw keys due to timing jitters of detectors and electronics. This experiment shows an efficient and convenient way to realize entanglement-based DO-QKD.
\par \setlength{\parskip}{0.1em}  The experimental setup of energy-time entanglement-based DO-QKD is shown schematically in Fig. \ref{fig1}. At Alice side, energy-time entangled photon pairs are generated through SFWM in a piece of silicon waveguide under CW pumping. The length of the waveguide is 5 mm and the wavelength of the pump light is 1549.32 nm. Then the signal and idler photons of each photon pairs are selected out by a filter system made of cascaded dense wavelength division multiplexer (DWDM). The central wavelengths of the generated signal and idle photons were 1546.12nm and 1552.52nm, respectively. Signal photons are used for Alice’s local measurement. The idler photons are sent to Bob through single mode fibers of 20 km. A dispersion compensation module (DCM) are introduced to compensate the dispersion introduced by the optical fibers. At each side, a 50:50 fiber coupler routes photons into two paths randomly. In one path, photons are detected directly, which corresponds to the time basis, and the time of single photon detection events are recorded, which are mainly used for the key generation. In the other path, photons are detected and recorded after a normal dispersion (ND, at Alice side) or abnormal dispersion (AD, at Bob side) module ($\pm$ 1800 ps/nm), which corresponds to the frequency basis and is used for security test. In the experiments, four NbN superconducting nanowire single-photon detectors (SNSPDs, fabricated by SIMIT, CAS, China) are used, with detection efficiency of ~50\% at 1550 nm. Their dark-count rates are about 100 counts per second, average timing jitters are ~80ps. Single photon events are precisely recorded by a time-to-digital converter (TCSPC, PicoQuant HydraHarp 400) under a resolution of 1 ps and then sent to a computer for data processing.
\par \setlength{\parskip}{0.1em} At Alice and Bob sides, photons are detected by either the time basis or frequency basis, and their time resolved coincidence are used for security test. The coincidence between the photons of time bases at both sides show a narrow peak. The full width at half maximum (FWHM) of the coincidence peak is mainly determined by the timing jitters of single photon detectors. The coincidence of the photons of frequency bases also shows a narrow peak, which is the effect of nonlocal dispersion cancellation introduced by the AD and ND components at Alice and Bob sides. On the contrary, if the coincidence measurement is taken between the photons of different bases at the two sides, the coincidence peak is broadened by the dispersive component at the frequency basis. By this way, the unbiased time-frequency bases are constructed by introducing the dispersive components. Based on the single photon events detected under frequency bases and a part of events detected under time bases, joint measurements of time-frequency covariance matrix (TFCM) can be carried out. It is used to evaluate Shannon information between Alice and Bob, and Eve’s maximum accessible information, which is responsible for the system security \cite{PhysRevA.87.062322}. The secure information that Alice and Bob could extract from per coincidence \cite{doi:10.1098/rspa.2004.1372,PhysRevLett.97.190503} is expressed as:
\begin{equation}
\label{e1}
\Delta I=\beta I(A;B)-\chi(A;E)
\end{equation}
where $\beta$ is the reconciliation efficiency, $I(A;B)$ is the Shannon information between Alice and Bob.The Eve’s maximum accessible information are quantified by the Holevo information $\chi(A;E)$.To evaluate $\Delta I$, the security analysis of DO-QKD follows the well-established proofs for protocols of Gaussian continuous-variable quantum key distribution (CV-QKD), which are based on the optimality of Eve’s Gaussian collective attack for a given TFCM \cite{PhysRevLett.97.190503,RevModPhys.84.621,0953-4075-37-2-L02}.Eve’s actions will disturb Alice and Bob’s initial TFCM. The excess spectral noise factor $\xi_w$ is used to quantify the disturbance towards the TFCM, which is expressed as:
\begin{equation}
\label{e2}
\xi_w=\frac{\sigma_{w}^{2}}{\sigma_{w_0}^{2}}-1
\end{equation}
where $\sigma_w^2$ quantifies the spectral correlation between detected photons at Alice and Bob, and $\sigma_{w_0}^2$ represents the noiseless correlation excluding excess channel noise or Eve's intrusion. In the practical measurement, we take the back-to-back experiment configuration with negligible channel loss as the noiseless correlation case.The corresponding correlation characteristics between Alice and Bob are treated as the noiseless correlation $\sigma_{w_0}^2$. Then, $I(A;B)$ and $\chi(A;E)$ can be calculated by the covariance matrix approach \cite{PhysRevA.87.062322,PhysRevLett.97.190503}, and $\Delta I$ can be estimated according to Eq. (\ref{e1}).
\begin{figure}[htbp]
\vspace{-0.1cm}
\setlength{\abovecaptionskip}{-0.1cm}
\setlength{\belowcaptionskip}{-0.1cm}  
\centering
\includegraphics[width=7.5cm]{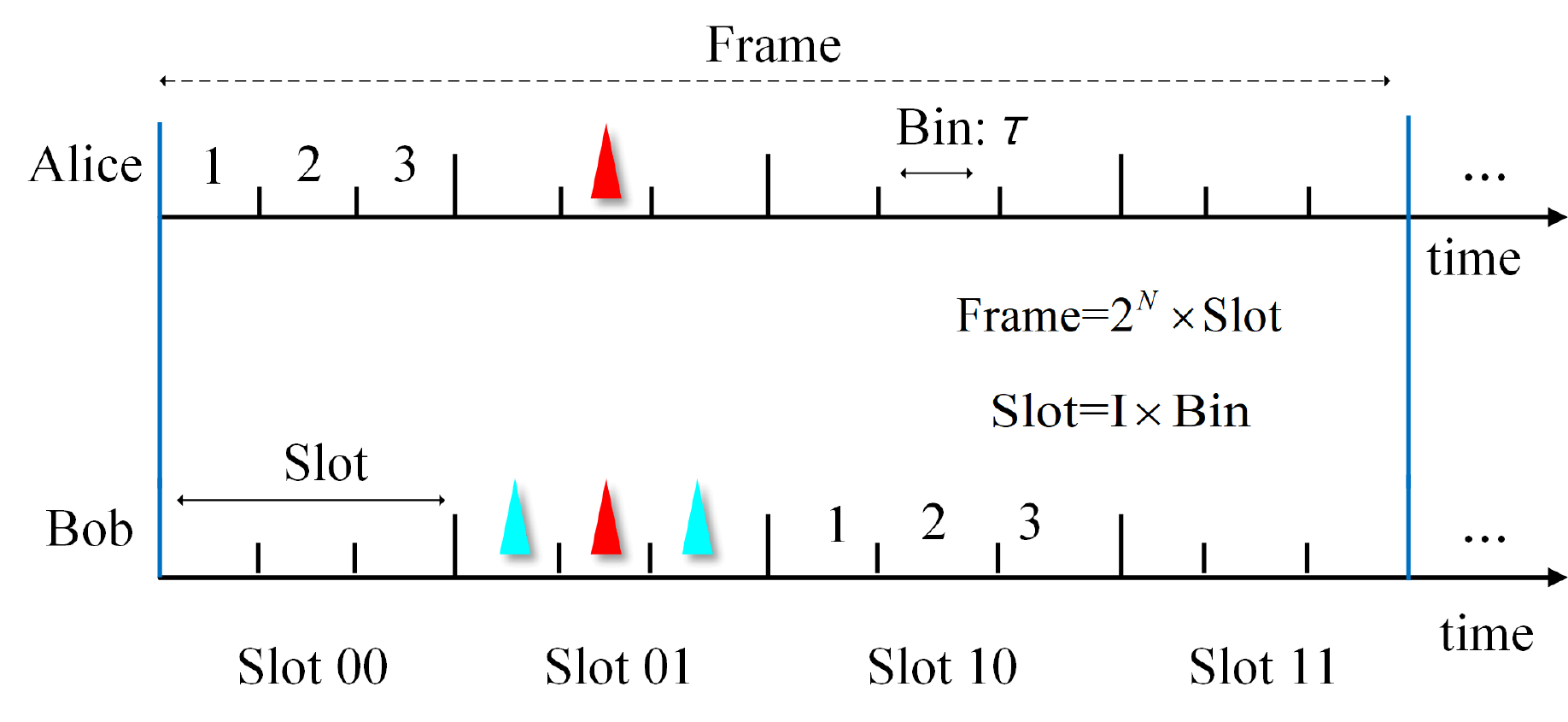}%
\caption{\label{fig2}The format of high dimensional encoding in the experiment. A frame has $M=2N$ slots and a slot has $I$ bins. In this example, $I=3$, $N=2$, $M=4$.}%
\end{figure}

Alice and Bob build their raw keys from correlated single photon events acquired in the time bases. The single photon events are sifted in a large alphabet way to increase the information per coincidence \cite{PhysRevLett.98.060503}, under a format of high dimensional encoding shown in Fig. \ref{fig2}. In the format, a time frame includes $M(M=2^N)$ consecutive slots and a slot including $I$ time bins. The Bin width is denoted by $\tau$. In the sifting process, firstly Alice and Bob communicate their frame numbers with one single photon detection event and keep the frames they both obtained one event. Then, for each kept frame, Alice and Bob check their time bin numbers which indicate the time bins they record their detection events. In Fig. \ref{fig2}, the two pulses with red color indicate the correlated detection events and the two pulses with light green color indicate the possible errors due to the non-zero coincidence peak width, which is mainly due to timing jitters of single photon detectors in this experiment setup. It can be expected that the checking process of time bin numbers can significantly reduce the QBER \cite{PhysRevLett.98.060503}. If the time bin numbers of Alice and Bob are the same, the frame is selected out in which the single photon detection events at Alice and Bob are considered as a coincidence event. Eventually, the raw keys are generated by the slot numbers of the coincidence events in the selected frames at Alice and Bob. Therefore, in this format one coincidence event generates raw keys of $N=log_{2}M$ bits. It can be seen that Alice and Bob do not directly communicate the information of slots and avoid revealing the slot numbers to Eve. The parameters of this bin sifting process should be optimized according to the system conditions, which determines the performance of key generation. After the bin sifting processes, raw keys are generated at Alice and Bob and then they proceed to perform error correction and privacy amplification to generate secret keys.
\begin{figure}[htbp]
\vspace{-0.1cm}
\setlength{\abovecaptionskip}{-0.1cm}
\setlength{\belowcaptionskip}{-0.1cm}  
\centering
\includegraphics[width=8cm]{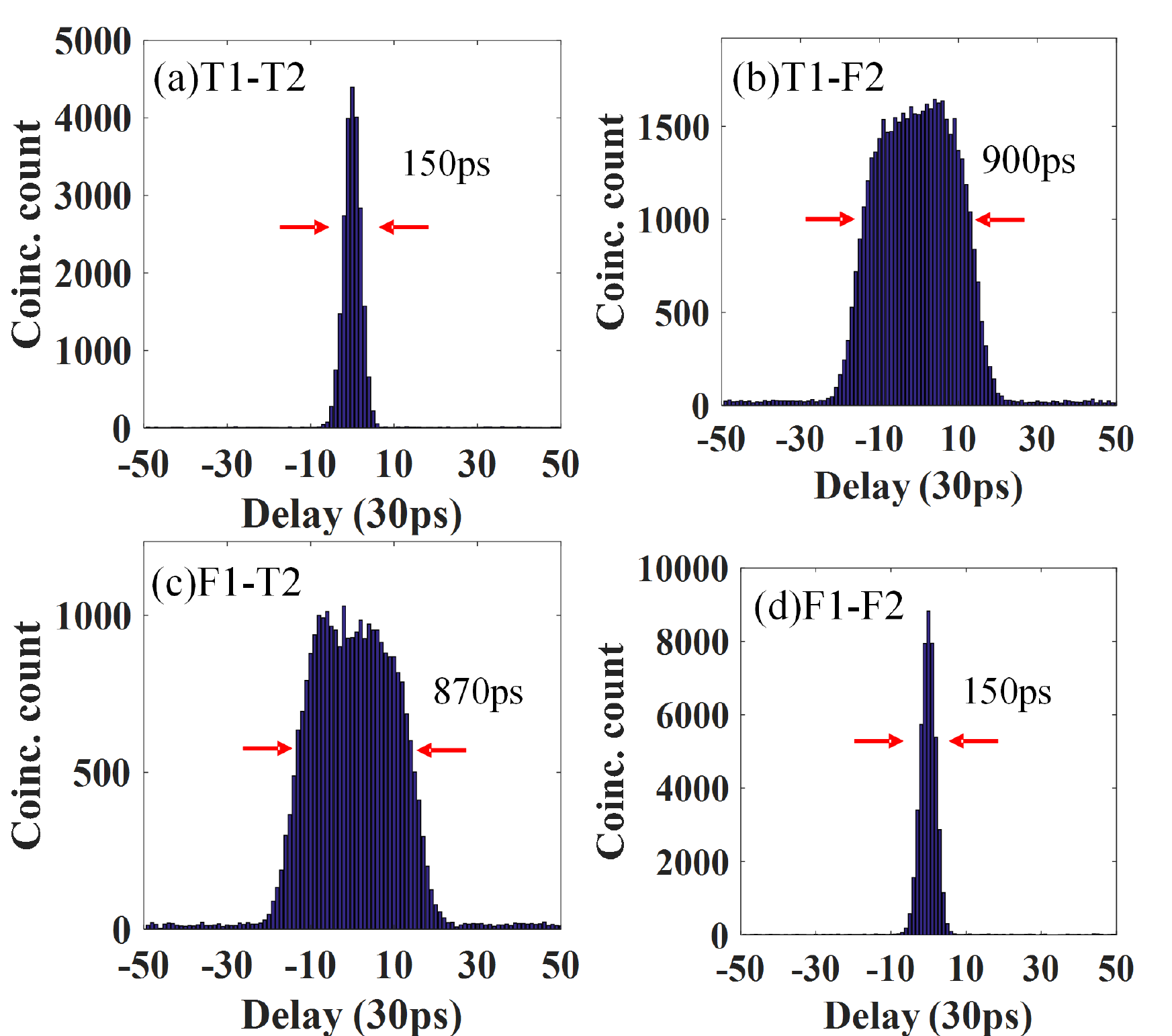}%
\caption{\label{fig3}Time resolved coincidence counts between Alice and Bob under four possible basis combinations. (a) Coincidence counts under T1 and T2; (b) Coincidence counts under T1 and F2; (c) Coincidence counts under F1 and T2; (d) Coincidence counts under F1 and F2; All the coincidence counts were acquired in 5s.}%
\end{figure}

As shown in Fig. \ref{fig1}, the time bases and frequency bases at Alice and Bob sides are marked as T1, F1, T2, and F2, respectively. The experiments were carried out with optical fiber transmission of 20 km under a pump level that the single photon count rates of SNSPDs of the four corresponding bases are 554kHz, 321kHz, 315kHz, and 245kHz, respectively. Fig. \ref{fig3} shows the time resolved coincidence counts between Alice and Bob under four possible basis combinations, which were acquired in 5s and used to take the calculation of security test of DO-QKD. The bin width for the coincidence measurement is 30ps. It should be noted that only 30\% of the single photon detection events of time bases at Alice and Bob were selected randomly to calculate the coincidences for the security test and rest of them were used to generate raw keys.

Fig. \ref{fig3}(a) is the coincidence counts of T1-T2. It can be seen that the coincidence peak of the single photon detection events recorded under time bases at both Alice and Bob is a narrow peak with a FWHM of ~150ps. The timing jitters of SNSPDs mainly contributes to the peak width. Fig. \ref{fig3}(b) and Fig. \ref{fig3}(c) are the coincidences of T1-F2 and F1-T2, i.e. the single photon detection events are recorded under different bases at Alice and Bob. In these two cases, the coincidence peaks are broadened to 900ps and 870 ps, respectively, by the dispersive components at the frequency bases. Fig. \ref{fig3}(d) is the coincidences of F1-F2. In this case, the single photon events were recorded under frequency bases at both sides. It can be seen that narrow coincidence peak recovers due to the nonlocal dispersion cancellation with a FWHM of 150ps. According to the experimental results shown in Fig. \ref{fig3}, the corresponding TFCM can be calculated. By further decomposing the TFCM, we obtained an upper bound on the Holevo information of $\chi(A;E)=0.211$ bpc (bit per coincidence), which indicates the impact of excess channel noise and possible disturbance of Eve introduced by the fiber transmission.
\begin{figure}[htbp]
\vspace{-0.1cm}
\setlength{\abovecaptionskip}{-0.1cm}
\setlength{\belowcaptionskip}{-0.1cm}  
\centering
\includegraphics[width=8.5cm]{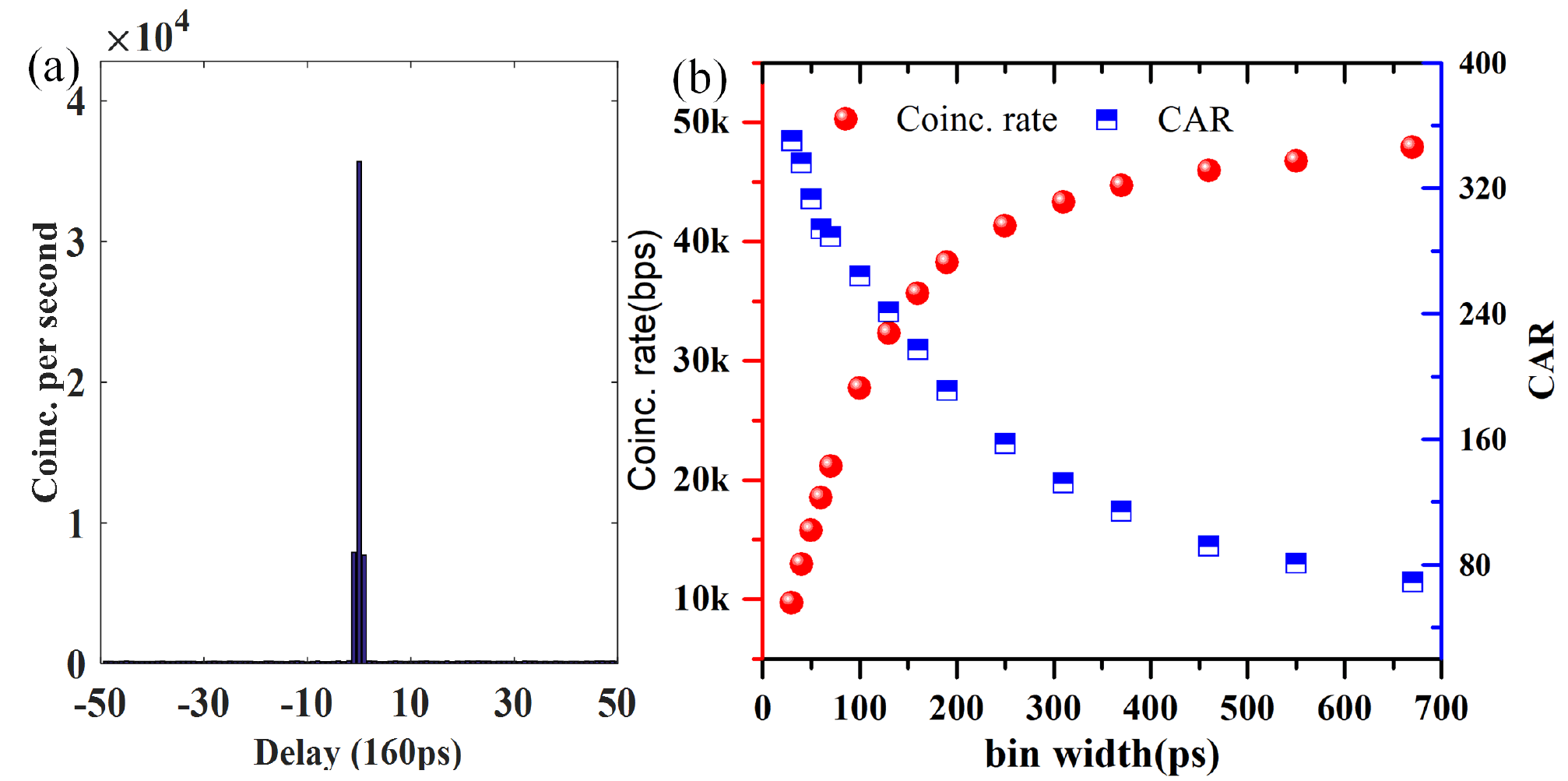}%
\caption{\label{fig4}The performance of the coincidence events for key generation. (a) A typical time resolved coincidence peak under a time bin of 160ps. (b) The coincidence rate and the coincidence to accidental coincidence ratio (CAR) under different time bin widths.}%
\end{figure}

For the key generation, Alice and Bob built their keys from 70\% of the single photon detection events acquired in the time bases at both side. Fig. \ref{fig4} shows the performances of the coincidence events for key generation. Fig. \ref{fig4}(a) shows a typical coincidence peak between the detection events that Alice and Bob recorded at time bases. The coincidence was recorded under a time bin of 160ps. The corresponding coincidence rate and CAR was 35.6KHz and 217, respectively. It can be seen that the central time bin in the coincidence peak has the maximum contribution on the coincidence. Here, we defined the coincidence count rate record in this time bin as the effective coincidence count rate, since only the coincidence events in this bin contribute to the key generation according to the bin sifting process. It can be expected that the reduction of the bin width would lead to the decrease of effective coincidence count rate if the bin width was smaller than the width of the coincidence peak. The effective CAR was defined as the ratio between the effective coincidence rate and the average accidental coincidence rate in bins outside the coincidence peak. Fig. \ref{fig4}(b) shows the effective coincidence count rate and effective CAR calculated under different time bin width. It can be seen that the effective coincidence rate rise obviously under increasing time bin width, which shows that the increase of bin width would improve the key generation rate. While, the effective CAR decreases under increasing bin width since the accidental coincidence count rate in one bin rises more rapidly. It can be expected that smaller bin width is preferred to reduce the QBER of the raw key. Hence, tradeoff should be taken on the bin width selection. 
\begin{figure}[htbp]
\vspace{-0.1cm}
\setlength{\abovecaptionskip}{-0.1cm}
\setlength{\belowcaptionskip}{-0.2cm}  
\centering
\includegraphics[width=7cm]{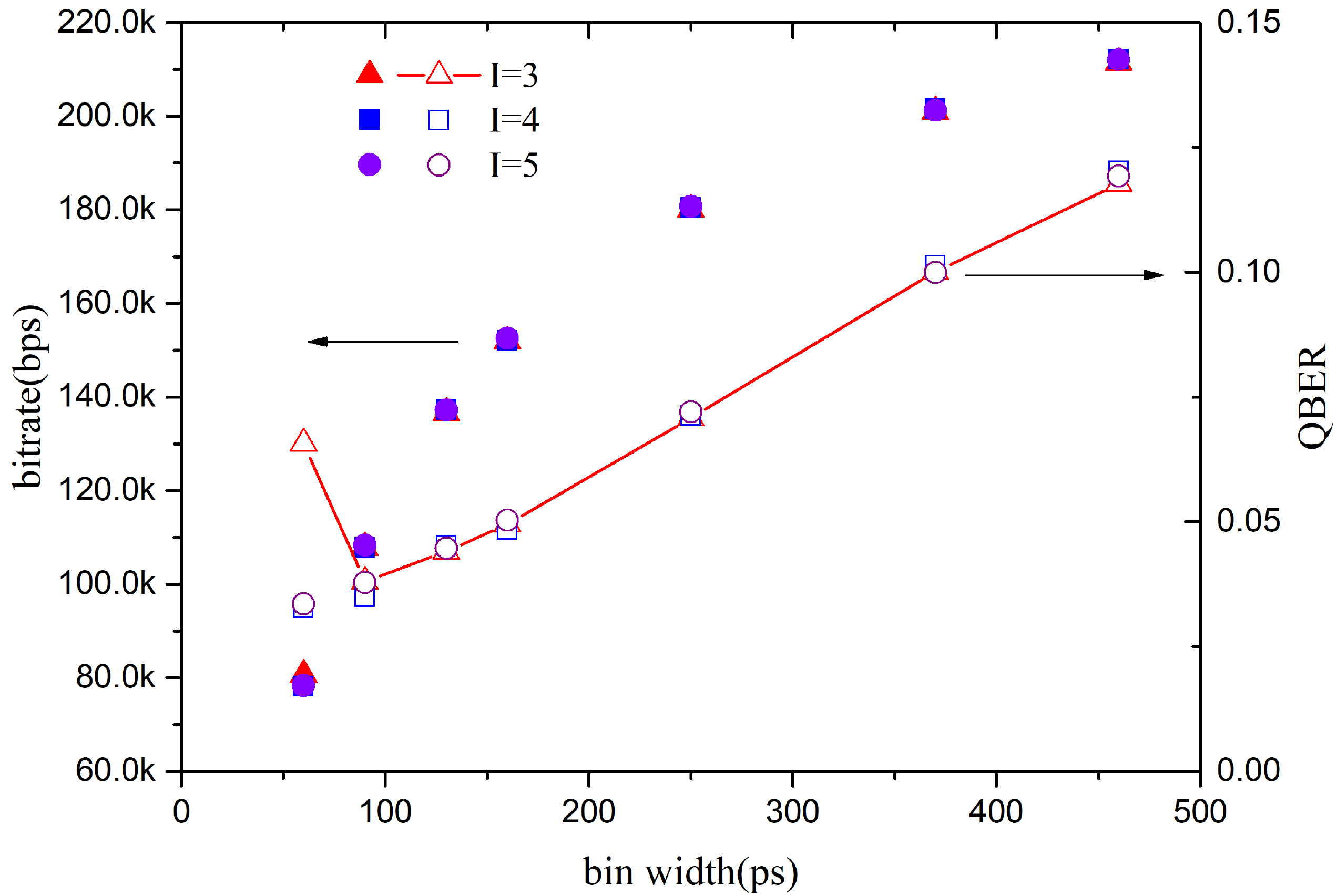}%
\caption{\label{fig5}Raw key generation rate and QBER under different bin width $\tau$ and division numbers $I$.}%
\end{figure}

To optimize the performance of the DO-QKD system in this work, we firstly considered a specific three level format, in which a frame has 16 slots, hence, $N=4$. The raw key generation rate and QBER were calculated according the experimental data under different bin width $\tau$ and bin number $I$ in a slot. The results are shown in Fig. \ref{fig5}. 

Fig. \ref{fig5} shows that the raw key generation rate increases with the increasing bin width, on the other hand, the QBER is also mainly increases with the increasing bin width under all the bin number $I$. These results can be explained by Fig. \ref{fig4}(b). However, in the case of $I=3$, the QBER increases with decreasing bin width when the bin width is small. The reason is that if the slot is too narrow to cover the coincidence peak, the coincidence located the same bin but adjacent slots at the two sides would introduce obvious errors. Hence, there is a bin width for the minimum QBER when $I=3$. It can be expected that in the cases of $I$= 4 and 5, the minimum QBERs also exist. However, they would appear at smaller bin widths which is not shown in Fig. \ref{fig5}, since they have lager slot widths if the bin width is fixed. To optimize the performance when $I=3$, we could choose appropriate parameters to achieve maximum raw key generation rate under a specific QBER requirement according to Fig. \ref{fig5}. For example, if the requirement is set as QBER $\leq$ 5\%, we could obtain the optimized parameters that the bin width $\tau$ is 160 ps and bin number in a slot $I$ is 3. In this case, the raw key generation rate is 151 kbps with a QBER of 4.95\%. 
\begin{figure}[htbp]
\vspace{-0.1cm}
\setlength{\abovecaptionskip}{-0.1cm}
\setlength{\belowcaptionskip}{-0.5cm}  
\centering
\includegraphics[width=7cm]{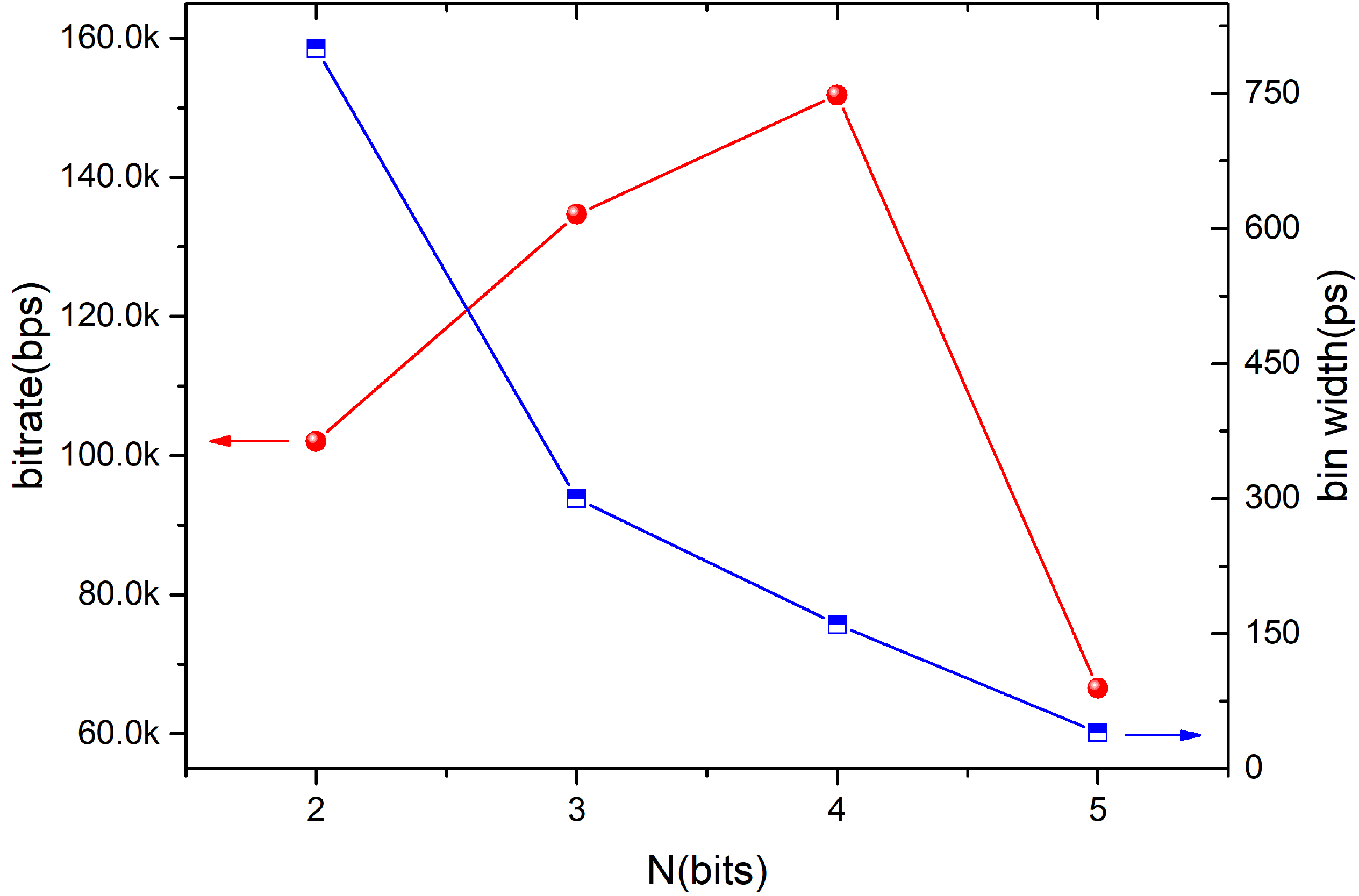}%
\caption{\label{fig6}Optimized bitrate and the corresponding bin widths under different dimension N with QBER all bounding below 5\%.}%
\end{figure}

Then, we calculated the system performance under different dimension $N$ and obtained the optimized format parameters and corresponding system performance under QBER $\leq$ 5\%. The results are shown in Fig. \ref{fig6}. It can be seen that the raw key generation rate reaches its maximum (151 kHz) when the dimension is $N=4$ with a bin width of 160ps. If $N$ is smaller than 4, the key generation per coincidence is limited. On the other hand, if the dimension $N$ increases to $N=5$, smaller bin width is required to guarantee QBER $\leq$ 5\%, which highly reduce the effective coincidence count rate. It is worth noting that in this work the QBER is highly restrained by the three level bin sifting process.

We optimized parameters of the format according to Fig. \ref{fig5} and Fig. \ref{fig6}. As a result, a raw key generation rate of 151kbps with a QBER of 4.95\% can be achieved under $N=4$, $I=3$ and $\tau$=160ps. Then, in this case secret keys were extracted from raw ones after error reconciliation and privacy amplification. Based on the acquired raw key with low QBER, the low-density parity-check (LDPC) code \cite{1057683} were adopted for error reconciliation, which was efficient at data interaction. The reconciliation efficiency $\beta=90\%$ was achieved at this low QBER (4.95\%). Privacy amplification was implemented using hash functions \cite{PhysRevLett.77.2818}. The secret key rate could be estimated by Eq. (\ref{e1}). The analysis of Fig. 3 has shown that Eve’s Holevo information $\chi(A;E)$ is 0.211 bpc, which is estimated by the calculated TFCM. The Shannon information between Alice and Bob $I(A;B)$ can be estimated by similar way, which is 3.48 bpc in this experimental system. As a result, the secret key capacity $\Delta I$ is 2.92 bpc according to Eq.(\ref{e1}), leading to a secret key rate of 104 kbps.

In this paper, we experimentally demonstrate the entanglement-based DO-QKD with high dimensional encoding over optical fibers of 20 km, in which the energy-time entangled photon pairs were generated by SFWM in a piece of silicon waveguide. A DCM is utilized to compensate the chromatic dispersion introduced by single mode fibers of 20 km. By three level bin sifting towards the single photon events measured based on the time bases, the QBER is significantly reduced to about 4.95\% under a raw key rate of 151 kbps. The whole system is secured by nonlocal dispersion cancellation based on dispersive optics. Finally, a secure key capacity of 2.92 bpc is achieved after subtracting Eve’s Holevo information. These experimental results show that entanglement based DO-QKD protocol can be implemented in an efficient and practicable way. It has great potential on quantum secure communication networks in the future. 

\vspace{0.3cm}
This work was supported by the National Key R\&D Program of China under Contract No. 2017YFA0303704 and No. 2017YFA0304000; the National Natural Science Foundation of China under Contract No. 61575102, No. 91750206, No. 61671438 and No. 61621064; Tsinghua National Laboratory for Information Science and Technology. 


\begin{thebibliography}{10}
	
	\bibitem{bennett1984quantum}
	H~Bennett~Ch and G~Brassard.
	\newblock Quantum cryptography: public key distribution and coin tossing int.
	\newblock In {\em Conf. on Computers, Systems and Signal Processing (Bangalore,
		India, Dec. 1984)}, pages 175--9, 1984.
	
	\bibitem{PhysRevLett.68.557}
	Charles~H. Bennett, Gilles Brassard, and N.~David Mermin.
	\newblock Quantum cryptography without bell's theorem.
	\newblock {\em Phys. Rev. Lett.}, 68:557--559, Feb 1992.
	
	\bibitem{PhysRevLett.94.230503}
	Xiang-Bin Wang.
	\newblock Beating the photon-number-splitting attack in practical quantum
	cryptography.
	\newblock {\em Phys. Rev. Lett.}, 94:230503, Jun 2005.
	
	\bibitem{doi:10.1063/1.2432296}
	Yi~Zhao, Bing Qi, and Hoi-Kwong Lo.
	\newblock Experimental quantum key distribution with active phase
	randomization.
	\newblock {\em Applied Physics Letters}, 90(4):044106, 2007.
	
	\bibitem{doi:10.1063/1.5027030}
	Alberto Boaron, Boris Korzh, Raphael Houlmann, Gianluca Boso, Davide Rusca,
	Stuart Gray, Ming-Jun Li, Daniel Nolan, Anthony Martin, and Hugo Zbinden.
	\newblock Simple 2.5 ghz time-bin quantum key distribution.
	\newblock {\em Applied Physics Letters}, 112(17):171108, 2018.
	
	\bibitem{PhysRevLett.98.010505}
	Cheng-Zhi Peng, Jun Zhang, Dong Yang, Wei-Bo Gao, Huai-Xin Ma, Hao Yin, He-Ping
	Zeng, Tao Yang, Xiang-Bin Wang, and Jian-Wei Pan.
	\newblock Experimental long-distance decoy-state quantum key distribution based
	on polarization encoding.
	\newblock {\em Phys. Rev. Lett.}, 98:010505, Jan 2007.
	
	\bibitem{Lucamarini:13}
	M.~Lucamarini, K.~A. Patel, J.~F. Dynes, B.~Fr\"{o}hlich, A.~W. Sharpe, A.~R.
	Dixon, Z.~L. Yuan, R.~V. Penty, and A.~J. Shields.
	\newblock Efficient decoy-state quantum key distribution with quantified
	security.
	\newblock {\em Opt. Express}, 21(21):24550--24565, Oct 2013.
	
	\bibitem{PhysRevA.76.012307}
	Xiongfeng Ma, Chi-Hang~Fred Fung, and Hoi-Kwong Lo.
	\newblock Quantum key distribution with entangled photon sources.
	\newblock {\em Phys. Rev. A}, 76:012307, Jul 2007.
	
	\bibitem{doi:10.1063/1.2348775}
	Ivan Marcikic, Antía Lamas-Linares, and Christian Kurtsiefer.
	\newblock Free-space quantum key distribution with entangled photons.
	\newblock {\em Applied Physics Letters}, 89(10):101122, 2006.
	
	\bibitem{PhysRevLett.81.5932}
	H.-J. Briegel, W.~D\"ur, J.~I. Cirac, and P.~Zoller.
	\newblock Quantum repeaters: The role of imperfect local operations in quantum
	communication.
	\newblock {\em Phys. Rev. Lett.}, 81:5932--5935, Dec 1998.
	
	\bibitem{PhysRevA.87.062322}
	Jacob Mower, Zheshen Zhang, Pierre Desjardins, Catherine Lee, Jeffrey~H.
	Shapiro, and Dirk Englund.
	\newblock High-dimensional quantum key distribution using dispersive optics.
	\newblock {\em Phys. Rev. A}, 87:062322, Jun 2013.
	
	\bibitem{lee2016high}
	Catherine Lee, Darius Bunandar, Zheshen Zhang, Gregory~R Steinbrecher, P~Ben
	Dixon, Franco~NC Wong, Jeffrey~H Shapiro, Scott~A Hamilton, and Dirk Englund.
	\newblock High-rate field demonstration of large-alphabet quantum key
	distribution.
	\newblock {\em arXiv preprint arXiv:1611.01139}, 2016.
	
	\bibitem{PhysRevA.90.062331}
	Catherine Lee, Zheshen Zhang, Gregory~R. Steinbrecher, Hongchao Zhou, Jacob
	Mower, Tian Zhong, Ligong Wang, Xiaolong Hu, Robert~D. Horansky, Varun~B.
	Verma, Adriana~E. Lita, Richard~P. Mirin, Francesco Marsili, Matthew~D. Shaw,
	Sae~Woo Nam, Gregory~W. Wornell, Franco N.~C. Wong, Jeffrey~H. Shapiro, and
	Dirk Englund.
	\newblock Entanglement-based quantum communication secured by nonlocal
	dispersion cancellation.
	\newblock {\em Phys. Rev. A}, 90:062331, Dec 2014.
	
	\bibitem{doi:10.1098/rspa.2004.1372}
	Igor Devetak and Andreas Winter.
	\newblock Distillation of secret key and entanglement from quantum states.
	\newblock {\em Proceedings of the Royal Society A: Mathematical, Physical and
		Engineering Sciences}, 461(2053):207--235, 2005.
	
	\bibitem{PhysRevLett.97.190503}
	Ra\'ul Garc\'{\i}a-Patr\'on and Nicolas~J. Cerf.
	\newblock Unconditional optimality of gaussian attacks against
	continuous-variable quantum key distribution.
	\newblock {\em Phys. Rev. Lett.}, 97:190503, Nov 2006.
	
	\bibitem{RevModPhys.84.621}
	Christian Weedbrook, Stefano Pirandola, Ra\'ul Garc\'{\i}a-Patr\'on, Nicolas~J.
	Cerf, Timothy~C. Ralph, Jeffrey~H. Shapiro, and Seth Lloyd.
	\newblock Gaussian quantum information.
	\newblock {\em Rev. Mod. Phys.}, 84:621--669, May 2012.
	
	\bibitem{0953-4075-37-2-L02}
	Alessio Serafini, Fabrizio Illuminati, and Silvio~De Siena.
	\newblock Symplectic invariants, entropic measures and correlations of gaussian
	states.
	\newblock {\em Journal of Physics B: Atomic, Molecular and Optical Physics},
	37(2):L21, 2004.
	
	\bibitem{PhysRevLett.98.060503}
	Irfan Ali-Khan, Curtis~J. Broadbent, and John~C. Howell.
	\newblock Large-alphabet quantum key distribution using energy-time entangled
	bipartite states.
	\newblock {\em Phys. Rev. Lett.}, 98:060503, Feb 2007.
	
	\bibitem{1057683}
	R.~Gallager.
	\newblock Low-density parity-check codes.
	\newblock {\em IRE Transactions on Information Theory}, 8(1):21--28, January
	1962.
	
	\bibitem{PhysRevLett.77.2818}
	David Deutsch, Artur Ekert, Richard Jozsa, Chiara Macchiavello, Sandu Popescu,
	and Anna Sanpera.
	\newblock Quantum privacy amplification and the security of quantum
	cryptography over noisy channels.
	\newblock {\em Phys. Rev. Lett.}, 77:2818--2821, Sep 1996.
	
\end{thebibliography}

\end{document}